\documentclass
[aps,prl,twocolumn,superscriptaddress,showpacs,floatfix]{revtex4-1}%
\usepackage{graphicx,amsmath,amssymb}
\usepackage{amsmath}
\usepackage{amsfonts}
\usepackage{amssymb}
\usepackage{graphicx}%
\usepackage[utf8]{inputenc}
\usepackage{textcomp}
\providecommand{\U}[1]{\protect\rule{.1in}{.1in}}
\begin{document}
\title{Spontaneous Transition to a Correlated Phase of Skyrmions Observed in Real Space}
\author{John N. Moore}
\affiliation{Department of Physics, Tohoku University, Sendai 980-8578, Japan}
\author{Hikaru Iwata}
\affiliation{Department of Physics, Tohoku University, Sendai 980-8578, Japan}
\author{Junichiro Hayakawa}
\affiliation{Department of Physics, Tohoku University, Sendai 980-8578, Japan}
\author{Takaaki Mano}
\affiliation{National Institute for Materials Science, Tsukuba, Ibaraki 305-0047, Japan}
\author{Takeshi Noda}
\affiliation{National Institute for Materials Science, Tsukuba, Ibaraki 305-0047, Japan}
\author{Naokazu Shibata}
\affiliation{Department of Physics, Tohoku University, Sendai 980-8578, Japan}
\author{Go Yusa}
\email{yusa@m.tohoku.ac.jp}
\affiliation{Department of Physics, Tohoku University, Sendai 980-8578, Japan}
\date{\today
}

\begin{abstract}
We conduct photoluminescence microscopy that is sensitive to both electron and nuclear spin polarization to investigate the changes that occur in the magnetic ordering in the vicinity of the first integer quantum Hall state in a GaAs 2D electron system (2DES). We observe a discontinuity in the electron spin polarization and nuclear spin longitudinal relaxation time which heralds a spontaneous transition to a phase of magnetic skyrmions. We image in real space the spin phase domains that coexist at this transition, and observe hysteresis in their formation as a function of the 2DES's chemical potential. Based on measurements in a tilted magnetic field orientation, we found that the transition is protected by an energy gap containing the Zeeman energy, and conclude that the skyrmions here have formed as an ensemble.
\end{abstract}

\maketitle

A two-dimensional (2D) electron system (2DES) becomes strongly correlated at low temperatures when a strong magnetic field normal to it $B_{\perp}$ creates Landau levels (LLs) with energy separation large enough that many-body Coulomb interactions dominate the physics. The fractional quantum Hall (QH) effect, in which a gap in the density of states (DOS) is opened at fractional values of the LL filling factor $\nu$, is one result of these correlations, as are the enhancement of screening, magnetization and the transport activation gap at integer $\nu$ \cite{Eisenstein,Pascher,Meinel,Nicholas,Usher,Schmeller}. Particular interest has been given to the first integer QH state in GaAs, which is complicated beyond a single-particle picture by the presence of magnetic skyrmions \cite{Barrett}. Skyrmions are vortex-like spin textures resulting from a competition between Zeeman and Coulomb interactions hypothesized to form when electrons are added to or removed from the ferromagnetic state at $\nu=1$ \cite{Sondhi}. Extensive evidence of skyrmions comes from observed reductions in electron polarization $P$ and nuclear spin longitudinal relaxation time $T_1$, which are both consequences of the in-plane spin components of skyrmions \cite{Aifer,Melinte,Khandelwal,Zhitomirsky,Plochocka,Piot,Hashimoto,Guan,Cote,Groshaus}. However, there is only a very limited picture of how skyrmions form and interact.

Here we conduct detailed measurements of $P$ and $T_1$ using spin-sensitive photoluminescence (PL) microscopy to study the critical conditions at which QH skyrmions form while visualizing their long-range behavior. An abrupt change in $P$ and $T_1$, which occurs as the chemical potential of the 2DES traverses the neighborhood of $\nu=1$, signals a discontinuous transition to a skyrmion-rich phase. The discontinuity of the transition is confirmed by the real-space imaging of coexisting spin phase domains which exhibit hysteresis. We also show that the transition is protected by a gap in the DOS containing the Zeeman energy $E_Z$, and argue that at high $B_{\perp}$ the creation of one skyrmion triggers the spontaneous formation of a correlated ensemble of skyrmions.
\begin{figure}[b]
	\par
	\begin{center}
		\includegraphics{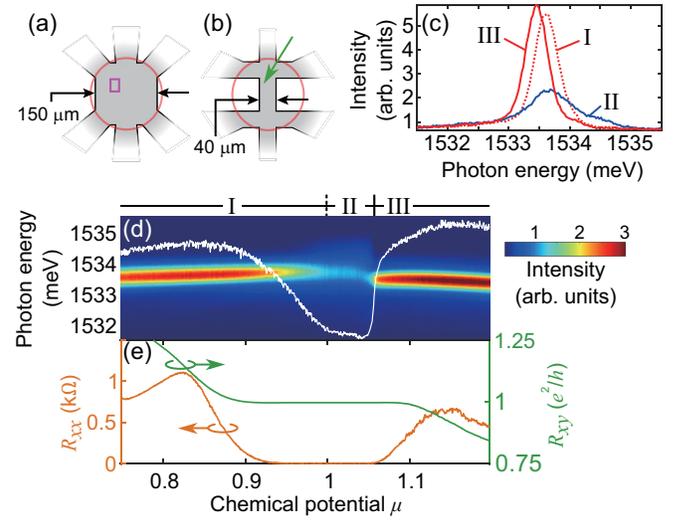}
	\end{center}
	\caption{Diagrams of Hall bars of (a) device A and (b) device B. Red circles: regions of illumination and macroscopic PL collection. Pink rectangle: region of PL spatial mapping. Green arrow: point of $T_1$ measurement ${\sim}6$ µm from edge. (c) PL spectra collected from regions I, II and III, specifically $\mu=0.86$, $1.03$ and $1.1$ respectively. (d)	PL spectra as a function of $\mu$ measured macroscopically. White curve: integrated intensity of the trion singlet state PL peak ("PL intensity"). (b) Longitudinal resistance $R_{xx}$ and Hall resistance $R_{xy}$ vs. $\mu$. Throughout, unless otherwise specified,  $T$ is ${\sim}45$ mK, $B$ is $8$ T perpendicular to the 2DES; $I_{sd}$ is $10$ nA, $13$ Hz.}%
	\label{fig:fig1}%
\end{figure}
\begin{figure*}[t]
	\par
	\begin{center}
		\includegraphics{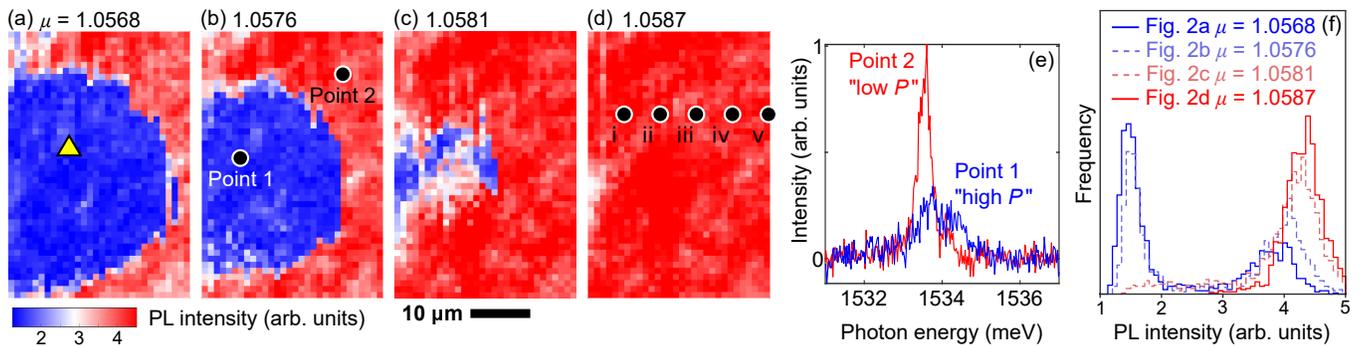}
	\end{center}
	\caption{(a)---(d) $33\times40$-$\mu$m$^{2}$ spatial images of the trion singlet state PL intensity at $\mu$ of: (a) $1.0568$; (b) $1.0576$; (c) $1.0581$; and (d) $1.0587$. (e) Example PL spectra collected microscopically in the high-$P$ region (blue) and low-$P$ region (red) as part of the spatial mapping. (f) PL intensity histograms of the images.}%
	\label{fig:fig2}%
\end{figure*}

Experiments were conducted in a dilution refrigerator on 2DESs in $15$-nm GaAs quantum wells (QWs) etched into a Hall bar geometry. We present results from two devices (devices A and B [Figs. 1(a) and 1(b)]) differing in electron mobility by ${\sim}50\%$ \cite{SI}. The devices are equipped with a back gate allowing electron density $n_e$ to be tuned by a gate voltage $V_g$ and allowing us to create the first integer QH state over a wide range of $B_{\perp}$ (up to $10$ T in device A), according to $\nu={\hbar}n_e/eB_{\perp}$, where $\hbar$ and $e$ are the reduced Plank constant and the elementary charge \cite{note4}. \textit{However, the exact value of $\nu$ is not accessible experimentally because it depends on the details of the macroscopic DOS under the influence of disorder and interactions.} In the literature conventionally, $n_e$ is approximated by $CV_g/e$, where $C$ is a constant capacitance per unit area between the 2DES and the gate electrode. More exactly, though, $C$ is not constant because the DOS has gaps, and strong electron interaction can cause the DOS to change unpredictably with small changes in $V_g$ and $B$. Therefore, $n_e$ and $\nu$ are not suitable parameters for our detailed examination of the small neighborhood around $\nu=1$. Instead, we define a unitless chemical potential $\mu\equiv\frac{hC_0V_{g}/e}{eB_{\perp}}+\alpha$; and $C_0V_{g}/e$ is the change in $n_e$ caused by $V_{g}$ when the DOS is constant, where $C_0$ is the gate's capacitance per unit area at total magnetic field $B\approx0$; $\alpha$ is an offset which we define such that $\mu=1$ at the center of the plateau in Hall resistance $R_{xy}$. Thus $\mu$ is comparable to the conventional estimate of $\nu$ in a gated 2DES.

We first study how $P$ changes in the QH liquid over the relatively wide range of $\mu=1~\pm {\sim}20\%$. Our method is to collect PL emitted from the QW, which has a spectrum sensitive to $P$ \cite{Hayakawa,MoorePRL}. We analyze the PL peaks from trions, which are created when a photoexcited hole and electron bind to an electron in the 2DES. Trions form in either a singlet state, in which the two electrons occupy opposite Zeeman levels (ZLs), or a triplet state, in which both electrons occupy the bottom ZL. Weak linearly polarized illumination excites the two trion states with rates sensitive to the occupation of the two ZLs in the 2DES. The singlet (triplet) PL intensity is thus anticorrelated (correlated) with $P$. 
\begin{figure}[t]
	\par
	\begin{center}
		\includegraphics{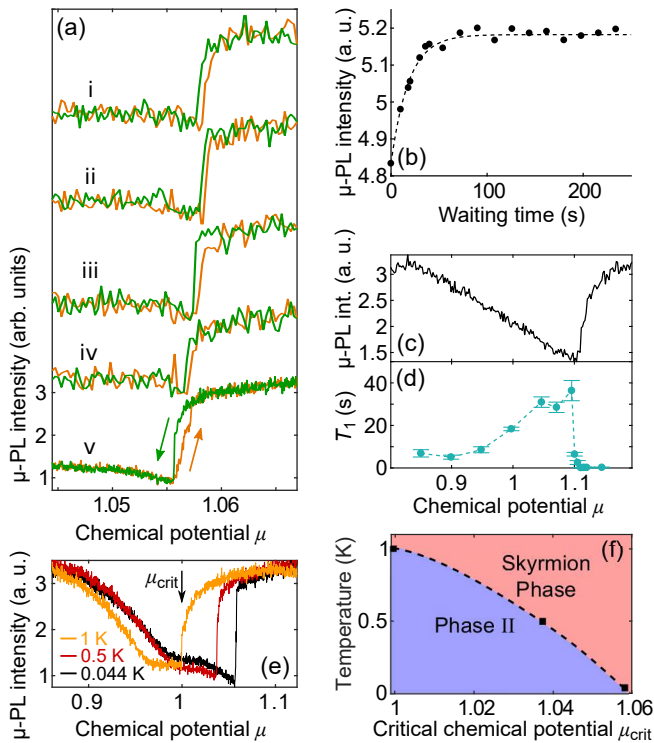}
	\end{center}
	\caption{(a) µ-PL intensity, offset for clarity, vs. $\mu$ at locations i-v measured by forward and backward scans of $\mu$ with round-trip scan times of $10.6$ min (i---iv) and $187.2$ min (v). (b) µ-PL intensity measured at $\mu\approx1/2$ vs. time of relaxation at $\mu=0.9974$. Dashed line: fitting function $a\exp{t}/{T_{1}}+b$, where $a=-0.3489$, $T_{1}=18.354$ s, and $b=5.1825$. (c) µ-PL intensity and (d) $T_{1}$ measured microscopically vs. $\mu$. In the experiments of (b)---(d), device B was used, and $T$ was ${\sim}40$ mK, $B=6$ T. (e) µ-PL intensity vs. $\mu$ at three temperatures. (f) Phase diagram in $T$ vs. $\mu_{\text{crit}}$. Dashed line is a guide to the eye. Error estimates are discussed in Supplemental Material \cite{SI}.}%
	\label{fig:fig3}%
\end{figure}

Figure 1(d) shows $\sigma^{-}$-polarized PL spectra, corresponding to the recombination of electrons from the bottom ZL, collected from a 165-µm diameter region (red circle) of the Hall bar in Fig. 1(a). PL spectra were measured simultaneously with transport [Fig. 1(e)] at $B$ of 8 T and $T$ of $45$ mK while scanning $\mu$ by increasing $V_g$. The PL peak appearing in red originates from the singlet state. After fitting it to a Lorentzian function, this peak’s intensity, hereafter called “PL intensity”, is integrated over a 390-µeV energy width and plotted as the white curve. Qualitatively we identify three regions of this curve, which we label as I, II and III. Figure 1(c) shows representative spectra from each region. In region II, the PL intensity shows a broad minimum, indicating a broad maximum in $P$, which has been observed in other studies and measured to have a value of $P\approx80\%$ \cite{Aifer,Zhitomirsky,Khandelwal,Zhitomirsky,Plochocka,Tiemann}. The ${\sim}20\%$ unpolarization means that the QH liquid is not perfectly ferromagnetic in region II; rather, it has been explained that a low density of disorder-localized skyrmions and/or antiskyrmions exists here \cite{Khandelwal,Zhitomirsky,Guan}. Experiments also conclude that in regions I and III where $P$ is greatly diminished, antiskyrmions and skyrmions are present as a crystal or liquid phase \cite{Khandelwal,Guan,Bayot,Gallais}. We now observe that while the transition between regions I and II is gradual and continuous, the transition between regions II and III is abrupt. 

We turn our attention to investigating the sharp transition in $P$ at $B=8$ T over the extremely narrow range $\mu=1.05775\pm0.09\%$. Collecting PL from a ${\sim}1$-µm spot (µ-PL), we performed scanning microscopy of PL intensity where the feature occurs, and obtained the images in Figs. 2(a)-2(d) taken inside the pink rectangle indicated in Fig. 1(a). We observed two distinct domains (blue and red) of differing PL intensity in Figs. 2(a)-2(b) to collapse into a homogeneous state of high PL intensity in Fig. 2(d). The strong contrast between the PL spectra in the two domains is apparent in the spectra in Fig. 2(e) obtained at points 1 and 2 of Fig. 2(b). The evolution of these images is also captured in their histograms [Fig. 2(f)], in which two well-separated peaks corresponding to the two domains give way to a single peak. We conclude that we have witnessed this area of the sample undergo a transition between the macroscopically identified regions II and III, corresponding to the blue and red domains respectively. We also find that whenever the two domains are both present, the autocorrelation coefficient for the images is larger than when $\mu\approx2/5$ \cite{SI}. This indicates that there is a mechanism of ordering as strong as the the QW's disorder potential (${\sim}100$ µeV \cite{Hayakawa}) which is causing the domains to extend over several tens of µm.

In Fig. 3(a) we present µ-PL intensity as a function of $\mu$ taken at the five points (i-v) indicated in Fig. 2(d), which have 6-µm spacing. Hysteresis appears when scanning $\mu$ across the transition. The hysteresis at location v was reproducible, having a width of $0.0006~\pm~0.0002$ \cite{SI}. We conclude from this hysteresis that the transition separating regions II and III in Fig. 1(d) is first-order.

We next seek to examine the magnetism of the 2DES from the perspective of $T_1$; $T_1$ is sensitive to the electronic spin ordering because hyperfine interaction with electrons is one mode of nuclear spin relaxation. Mismatch in electron and nuclear Zeeman energies normally makes this mode very inefficient, but if the mismatch is compensated by low-energy features of the electronic energy spectrum, a shortening of the nuclear spin relaxation rate occurs. $T_1$ is experimentally accessible to us because the PL intensity is also sensitive to the degree of local nuclear polarization $P_\text{N}$ by a mechanism we speculate is related to the Overhauser field’s influence on the trions \cite{MoorePRL,Abragam,Coish,note2}. In the following experiment, we dynamically create $P_\text{N}$ using the flip-flop scattering that occurs between electron and nuclear spins at the boundary of nonequilibrium striped spin phase domains at $\nu\approx2/3$ \cite{MoorePRL,MoorePRB}. This allowed us to detect the decay of $P_\text{N}$ in time by a procedure of repeatedly pumping the source-drain current $I_{sd}$ to create $P_\text{N}$ and then waiting a variable time before measuring the remaining $P_\text{N}$ by the PL intensity \cite{SI}. $T_1$ changes depending on the $\mu$ at which we wait.

Figure 3(b) shows the µ-PL intensity collected at the point in device B indicated by the green arrow in Fig. 1(b) as a function of the waiting time after pumping. Fitting these data to an exponential decay yields $T_{1}$. In Figs. 3(c) and 3(d) we show µ-PL intensity and $T_{1}$ measured at the same location as a function of $\mu$. Both of these data indicate a spontaneous change in spin ordering at roughly the same $\mu\approx1.1$ corresponding to the first-order transition \cite{note5}. We define this critical $\mu$ as $\mu_{\text{crit}}$. The extremely short value of $T_{1}$ above $\mu_{\text{crit}}$ and around $0.9$ in Fig. 3(d) is the expression of the Goldstone spin wave mode that is enabled by the in-plane U(1) degree of freedom of the spins within skyrmions and antiskyrmions \cite{Cote}. Thus, we confirm that region III contains a skyrmion-rich phase, and that region I contains antiskyrmions.

We state here, and confirm below, that region II corresponds to the gap in the DOS between the ZLs of the lowest LL. This gap should contain both Zeeman and Coulomb components. $\nu$ becomes approximately 1 at the bottom of this gap, and then does not increase significantly until the top of the gap, which is why $P$ has been observed to become plateau-like. At $\mu_{\text{crit}}$, $\nu$ increases sharply as skyrmions form up to some critical density. The discontinuity in $n_e$ across the transition might account for why the domain wall in Figs. 2(a) and 2(b) is concave with respect to phase II; the blue (red) region tends to expand (contract) in some places to maintain the low (high) density of phase II (the skyrmion phase). 

To investigate the energy gap, we measured µ-PL intensity vs. $\mu$ at three different temperatures [Fig. 3(e)] at the point indicated by the yellow triangle in Fig. 2(a). Increasing $T$ from $44$ mK to $1$ K caused $\mu_{\text{crit}}$ to shift to lower $\mu$, as is depicted in Fig. 3(f), and caused region II seen in the µ-PL intensity to shrink by ${\sim}50\%$ \cite{note6}. A combination of factors may be causing region II to shrink. Thermal energy might be exciting ground electrons at the Fermi level into the upper ZL, which then triggers skyrmions to form in a chain reaction. Additionally, the gap energy may be reduced when thermal fluctuations disrupt the exchange correlations stabilizing the spin-polarized regions of phase II. At higher $T$, the change in µ-PL intensity at $\mu_{\text{crit}}$ becomes diminished, possibly because at lower $\mu$ there is less chemical potential available for the skyrmion creation, or because skyrmions are reduced in size by thermal fluctuations.

The strong sensitivity of $\mu_{\text{crit}}$ to $T$ indicates that the gap energy is comparable to the thermal energy. We estimate the size of the gap's energy $E_g$ by noticing that region II appears to be shrinking linearly with $T$ such that it will disappear at $1.9\pm0.4$ K, which we take to be $E_g$ at the limit of $0$ K \cite{SI}. This is at most $\sim30\%$ larger than $E_Z$. This is consistent with either the Zeeman plus exchange energy cost of a single spin flip, or the cost of a skyrmion; the latter energy cost contains a large Zeeman component, but can have a net negative Coulomb component due to quantum fluctuations afforded by the superpositional states of the skyrmion's spins \cite{Fertig}.

We now probe the energy gap further by returning to measurements of PL intensity from the circular macroscopic region in Fig. 1(a). In order to isolate the contribution of Zeeman energy to the gap, we also tilt the device with respect to $B$ in the configuration of Fig. 4(a) while holding $B_{\perp}$ fixed. This increases $E_Z$ while leaving the orbital dynamics unchanged \cite{Fang}. Figure 4(b) shows PL intensity at tilt angels $\theta$ of $0^\circ$ (black) and $30^\circ$ (red) as a function of $\mu$ over a range of $B_{\perp}$. $\mu_{\text{crit}}$ is plotted in Fig. 4(c) against $B_{\perp}$ for both conditions of $\theta$. When tilting the sample $30^\circ$, the transition remains sharp, and $\mu_{\text{crit}}$ increases while region II tends to become wider. This indicates that the gap to the spontaneous formation of skyrmions has been expanded by the increase in $E_Z$. To measure the gap's expansion, we convert the difference in $\mu_{\text{crit}}$ between the two tilt angles $\Delta\mu_{\text{crit}}$ to units of kelvin based on the size of the gap measured at 8 T \cite{SI}. $\Delta\mu_{\text{crit}}$, plotted in Fig. 4(d), tends to scale with $B$, which confirms that it is proportional to $E_Z$; thus $E_g$ is linear with $E_Z$.

As $B_{\perp}$ grows, the PL intensity change vs. $\mu$ (or $V_{g}$) at the transition becomes sharper. This sharpness implies that upon creation of the first skyrmion, it takes very little additional chemical potential to create additional skyrmions up to a critical density. The additional chemical potential cost even seems to be negative because, as observed, the skyrmion phase spontaneously overcomes the background potential. This is nonintuitive, though, because skyrmions repel each other through short-range magnetic and long-range electrostatic interactions \cite{Timm}.

One explanation is based on the fact that adding or subtracting a skyrmion requires spin reorientations involving exchange of angular momentum with the environment; this process cannot occur as a single quantum transition. Only after the Fermi energy $E_F$ becomes high enough to excite a single spin-flipped electron into the spin-polarized 2DES, a skyrmion can begin to form with this electron as its nucleus. The surrounding spins relax into the configuration of a skyrmion, and the energy gained in this relaxation pays the cost of bringing another spin-flipped electron into the 2DES, which nucleates another skyrmion; thus skyrmions populate the 2DES in a chain reaction. The reverse of this process occurs when $E_F$ falls below the skyrmion ground state energy. This allows a skyrmion to be removed by extracting the electron at its nucleus from the skyrmion phase. The surrounding spins relax into a polarized configuration, which releases the energy that in turn expels further skyrmion nuclei from the 2DES in another chain reaction. In this view, the hysteresis of the transition derives from the difference between the single-spin flip quasiparticle and skyrmion ground state energies. 

In conclusion, here, the lowest-energy stable excitation above $\nu=1$ is not an individual skyrmion; it is a correlated spin texture containing an ensemble of skyrmions. This is a many-body effect which points to the mechanisms of skyrmion formation and interaction. 
\begin{figure}[t]
	\par
	\begin{center}
		\includegraphics{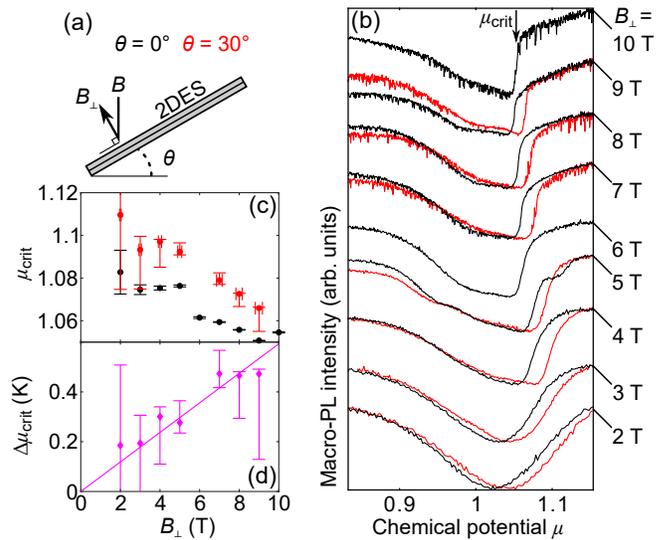}
	\end{center}
	\caption{(a) Schematic of the 2DES of device A tilted in the magnetic field. (b) Macroscopic PL intensity, normalized and offset for clarity, vs. $\mu$ as a function of $B_{\perp}$ when $\theta$=$0^\circ$ (black) and $30^\circ$ (red). (c) $\nu_{\text{crit}}$ vs. $B_{\perp}$ for $\theta=0^\circ$ (black) and $30^\circ$ (red). (d) $\Delta\mu_{\text{crit}}$ vs. $B_{\perp}$. Dashed line: fitting function $mB_{\perp}$, where $m=0.06\pm0.01$ K/T. Error estimates are discussed in Supplementary Material \cite{SI}.}%
\end{figure}

\begin{acknowledgments}
This work was supported by a Grant-in-Aid for Scientific Research (no.17H01037) from the Ministry of
Education, Culture, Sports, Science, and Technology (MEXT), Japan and the Asahi Glass Foundation. J.N.M. was
supported by a Grant-in-Aid from JSPS, MEXT and the Marubun Research Promotion
Foundation. J. H. was supported by a Grant-in-Aid from the Tohoku University
International Advanced Research and Education Organization.
\end{acknowledgments}


\begin{thebibliography}{99}
\bibitem {note2}$P_\text{N}$ can be ruled out as the cause of any of the phenomena so far discussed because the measurements do not employ high currents or intense circularly polarized illumination needed to dynamically generate $P_\text{N}$.
\bibitem {note4}$n_e$ is taken to be the average electron density over the area of the Hall bar. The local electron density at any given point is in general different from $n_e$.
\bibitem {note5}The jump in $T_{1}$ occurs at ${\sim}0.02$ lower $\mu$ than the jump in PL intensity. This is likely because at a location near to the measurement point the transition occurs at a lower $\mu$, and nuclear spin diffusion into that location enhances the effective relaxation rate.
\bibitem {note6}In the macro-PL intensity as well, we observe all of the same general changes with $T$ as seen microscopically.
\bibitem {note7}In device B, the continuous transition between phases I and II extends all the way up to $\mu_{\text{crit}}$; it is therefore difficult to evaluate the energy gap in region II quantitatively.
\end{thebibliography}
\end{document}